\documentclass[aps,pra,twocolumn,showpacs,superscriptaddress]{revtex4}
\usepackage[dvips]{epsfig}
\usepackage{graphicx}
\usepackage{amsfonts}
\usepackage{bm}
\usepackage{amsmath,amssymb,lscape,float}

\newcommand{\ket}[1]{{\vert #1 \rangle}}
\newcommand{\bra}[1]{{\langle #1 \vert}}

\newcommand{\abs}[1]{\left| #1 \right|} % for absolute value
\makeatletter
\renewcommand\appendix{\par
  \setcounter{section}{0}  \setcounter{subsection}{0}  \setcounter{equation}{0}  \gdef\thesection{\Alph{section}}
  \@addtoreset {equation}{section}
  \renewcommand{\theequation}{\thesection\arabic{equation}}}
\makeatother

\begin{document}
\title{Superfluid drag of two-species Bose-Einstein condensates in optical lattices}
\author{Patrick P. Hofer}
\affiliation{Department of Physics, University of Basel, Klingelbergstrasse 82, CH-4056
Basel, Switzerland}\author{C. Bruder}
\affiliation{Department of Physics, University of Basel, Klingelbergstrasse 82, CH-4056
Basel, Switzerland}
\author{Vladimir M. Stojanovi\'c}
\email{vladimir.stojanovic@unibas.ch}
\affiliation{Department of Physics, University of Basel, Klingelbergstrasse 82, CH-4056
Basel, Switzerland}
\date{\today}

\begin{abstract}
We study two-species Bose-Einstein condensates in quasi two-dimensional optical lattices of varying geometry and potential depth. Based on the numerically exact Bloch and Wannier functions obtained using
the plane-wave expansion method, we quantify the drag (entrainment coupling) between the condensate components. This drag originates from the (short range) interspecies
interaction and increases with the kinetic energy. As a result of the interplay between interaction and kinetic energy effects, the superfluid-drag coefficient shows a non-monotonic dependence on the lattice depth.
To make contact with future experiments, we quantitatively investigate the drag for mass ratios corresponding to relevant atomic species. 
\end{abstract}

\pacs{67.85.Fg, 67.85.Hj, 67.85.De}
\maketitle

\section{Introduction}
Owing to their many favorable features such as the absence of defects and dynamic control of relevant parameters,
optical lattices constitute a versatile tool for controlling the properties of atomic quantum fluids \cite{qfreviews}.
They have proven to be an invaluable platform for studying quantum phase transitions such as from a superfluid to a Mott-insulator (SF-MI) \cite{greiner_sfmi,bruder}
and simulating many-body systems \cite{lewenstein_rev}.
The high densities and low temperatures required for investigating these quantum effects are provided by Bose-Einstein condensates (BECs) \cite{leggett_rev,blakiebec,Liu+Wu:06}.

While multi-component BECs -- and, especially, two-component ones \cite{emuller,Chen+Wu:03,timmermans:03} -- have attracted a great deal of interest since the
early days of BEC \cite{bec_cornell,Davis+Ketterle:BEC:95,bec_3},  the addition of the optical-lattice environment \cite{bruder,bboptlatt,bec_twocomp} allows one to study interesting transport properties of 
cold atoms in the superfluid regime.
One of them, which has so far not been given due attention, is the superfluid drag in two-component BECs. Such a non-dissipative drag effect was first investigated by Andreev and Bashkin
\cite{superfluid_drag} in the context of $^3$He-$^4$He mixtures.
A microscopic theory of the drag between two weakly interacting Bose gases was developed in the
continuum limit \cite{continuum_limit} and generalized to the non Galilean invariant case as realized in an optical lattice \cite{linder_drag}.
For a system of strongly interacting bosons, the superfluid drag was investigated by means of Monte Carlo simulations \cite{negative_drag}.
Drag effects have also been studied in different systems. For instance, in electronic mesoscopic systems Coulomb drag has been studied both
theoretically and experimentally \cite{coulomb_drag}. 

In this work, we study the drag (entrainment coupling \cite{entrainment}) between two weakly interacting Bose gases in quasi two-dimensional optical lattices.
We derive a formula for the superfluid-drag coefficient valid for an arbitrary lattice and evaluate it for different lattice geometries. In this manner we extend the results presented in
Ref.~\onlinecite{linder_drag} where this has been done for the special case of the three-dimensional cubic lattice. 
This derivation proceeds by diagonalizing the Hamiltonian using a Bogoliubov approximation and subsequent expansion of the free energy in the superfluid velocities.

The band dispersions and interaction parameters needed to evaluate the superfluid-drag coefficient are obtained
numerically by means of a truncated plane-wave expansion of the optical lattice potentials.
Using this numerically exact approach, instead of the tight-binding approximation, we obtain results that are reliable even in the limit
of shallow optical lattices where superfluidity is assured for both commensurate and incommensurate filling. As a result of the interplay between the interaction and kinetic energies,
we find a non-monotonic dependence of the superfluid-drag coefficient on the optical lattice depth.
In contrast to Ref.~\onlinecite{linder_drag}, where the mass ratio which maximizes the superfluid drag was found to be around unity for an arbitrary lattice depth,
we find that the optimal mass ratio depends on the lattice depth.

In addition to the two-dimensional square lattice, we investigate the drag in the particularly interesting three-beam lattices (3BL).
These are two-dimensional optical lattices with non-separable
potentials \cite{double_well}, created by three in-plane laser beams \cite{blakie:04}. One special case of 3BLs is the triangular optical lattice,
which has lately received attention in connection with the experimental observation of the SF-MI transition \cite{sengstock_triangular}.

The paper is organized as follows. In Sec.~\ref{sec:sysmod} we present the theoretical description of two-species BECs in optical lattices.
At the same time we introduce the notations and conventions to be used throughout the paper.
We discuss the two-species Bose-Hubbard model in Sec.~\ref{sec:sysmoda}; Sec.~\ref{sec:sysmodb} is devoted to the numerical derivation of the parameters of the model, while
in Sec.~\ref{sec:sysmodc} we specify the optical-lattice potentials under investigation.\\
In Sec.~\ref{sec:drag} we generalize the derivation of an expression for the superfluid-drag coefficient at zero temperature, valid for an arbitrary lattice geometry.
The results for the superfluid drag in two different 3BLs and the four-beam square lattice are presented and discussed in Sec.~\ref{sec:results}. Finally, we conclude in Sec.~\ref{sec:conclusion}.

\section{System and model}
\label{sec:sysmod}
In this section, we describe the model of a two-species BEC in a quasi two-dimensional optical lattice.  We introduce the Hamiltonian of the system
(Sec.~\ref{sec:sysmoda}), outline the numerical evaluation of its parameters (Sec.~\ref{sec:sysmodb}) and discuss the two-dimensional optical lattices
under consideration (Sec.~\ref{sec:sysmodc}).
For convenience, we set $\hbar=1$ throughout the paper.

\subsection{Two-species Bose-Hubbard model}
\label{sec:sysmoda}
The single-band, two-species Bose-Hubbard model with short-range on-site interactions reads \cite{bruder}
\begin{equation}
\label{eq:hambh}
H=\sum_{ij,\alpha}\varepsilon_{ij}^{\alpha}b_{i\alpha}^{\dag}b_{j\alpha}
+\frac{1}{2}\sum_{i,\alpha\beta}U_{\alpha\beta}b_{i\alpha}^{\dag}b_{i\beta}^{\dag}b_{i\beta}b_{i\alpha}\:.
\end{equation}
The operator $b_{i\alpha}^{\dag}$ ($b_{i\alpha}$) creates (annihilates) a boson of component $\alpha=A,B$ occupying a Wannier orbital centered at lattice site $\boldsymbol{R}_i\;(i=1,...,N)$. In terms of the (three-dimensional) Wannier functions
$W_{\alpha}(\boldsymbol{r}),$
\begin{eqnarray} 
\label{eq:hambhpara}
\nonumber
&&\varepsilon_{ij}^{\alpha}=\int d^3r\:W^*_{\alpha}(\boldsymbol{r}-\boldsymbol{R}_i)\left[-\frac{\nabla^2}{2m_{\alpha}}+V(\boldsymbol{r})\right]W_{\alpha}(\boldsymbol{r}-\boldsymbol{R}_j),\\
&&U_{\alpha\beta}=\gamma_{\alpha\beta}\int d^3r\:\abs{W_{\alpha}(\boldsymbol{r})}^2\abs{W_{\beta}(\boldsymbol{r})}^2,
\end{eqnarray}
where the diagonal and off-diagonal elements of $\varepsilon_{ij}^{\alpha}$ respectively correspond to the on-site energies and the hopping amplitudes.
The potential $V(\boldsymbol{r})$ includes the lattice potential as well as a confinement in the $z$-direction; $\gamma_{\alpha\beta}\equiv2\pi(m_{\alpha}+m_{\beta})a_{\alpha\beta}/(m_{\alpha}m_{\beta})$
is determined by the particle masses and the $s$-wave scattering length $a_{\alpha\beta}$.

We consider a one-dimensional optical lattice in the $z$-direction, which provides the aforementioned confinement.
Approximating the confining potential around one of the minima leads to a harmonic potential with mass-dependent frequency $\omega_{z\alpha}$. As long as the corresponding oscillator
length $l_z\equiv(m\omega_{z\alpha})^{-1/2}$ is small compared to the scattering lengths, we can assume that the two-body collisions are not affected by the confinement and Eq. \eqref{eq:hambhpara}
gives the correct interaction parameters. For a sufficiently deep lattice in the $z$-direction, the Wannier functions can be written as
\begin{equation}
 \label{eq:wanfac}
W_{\alpha}(\boldsymbol{r},z)=\widetilde{W}_{\alpha}(\boldsymbol{r})\times\left(\frac{m_{\alpha}\omega_{z\alpha}}{\pi}\right)^{1/4}e^{-\frac{m_{\alpha}\omega_{z\alpha}}{2}z^2},
\end{equation}
where
$\boldsymbol{r}$ is a two-dimensional vector. Hereafter, all the bold letters denote two-dimensional vectors.

The $z$-dependence of $U_{\alpha\beta}$ can then easily be integrated out and we are left with
\begin{eqnarray}
 \label{eq:u2d}
 \nonumber
U_{\alpha\beta}=&\gamma_{\alpha\beta}&\sqrt{\frac{m_{\alpha}\omega_{z\alpha}m_{\beta}\omega_{z\beta}}{\pi(m_{\alpha}\omega_{z\alpha}+m_{\beta}\omega_{z\beta})}}\\*
&&\times\int d^2r\abs{\widetilde{W}_{\alpha}(\boldsymbol{r})}^2\abs{\widetilde{W}_{\beta}(\boldsymbol{r})}^2.
\end{eqnarray}

To switch from a real-space (lattice) description to momentum space, we use the Fourier-transformed boson operators
$b_{i\alpha}=N^{-1/2}\sum_{\boldsymbol{k}}b_{\boldsymbol{k}\alpha}e^{-i\boldsymbol{k}\boldsymbol{R}_i}$. Inserting this relation into Eq. \eqref{eq:hambh} leads to
\begin{eqnarray}
 \label{eq:hamk}
H&=&\sum_{\boldsymbol{k},\alpha}\varepsilon_{\boldsymbol{k}\alpha}b^{\dag}_{\boldsymbol{k}\alpha}b_{\boldsymbol{k}\alpha}\\*\nonumber
&&+\frac{1}{2N}\sum_{\alpha\beta}U_{\alpha\beta}\sum_{\boldsymbol{k}_1,...,\boldsymbol{k}_4}b^{\dag}_{\boldsymbol{k}_1\alpha}b^{\dag}_{\boldsymbol{k}_2\beta}b_{\boldsymbol{k}_3\beta}b_{\boldsymbol{k}_4\alpha}
\delta_{\boldsymbol{k}_1+\boldsymbol{k}_2,\boldsymbol{k}_3+\boldsymbol{k}_4}.
\end{eqnarray}
The band dispersion $\varepsilon_{\boldsymbol{k}\alpha}$ and the interaction parameters $U_{\alpha\beta}$ are calculated numerically as described below.

\subsection{Plane-wave expansion}
\label{sec:sysmodb}
Using the expansion of the Bloch functions in the reciprocal-lattice vectors $\boldsymbol{G}$ ($\mathcal{V}$ denotes the system volume)
\begin{equation}
 \label{eq:bloch}
\Psi_{\boldsymbol{k}\alpha}(\boldsymbol{r})=\frac{1}{\sqrt{\mathcal{V}}}\sum_{\boldsymbol{G}}C_{\boldsymbol{k},\boldsymbol{G}}^{\alpha}\:e^{i(\boldsymbol{k}+\boldsymbol{G})\cdot\boldsymbol{r}},
\end{equation}
the Bloch eigenvalue problem can be recast as
\begin{equation}
 \label{eq:eigenvalue}
\sum_{\boldsymbol{G}'}\bra{\boldsymbol{k}+\boldsymbol{G}}H_1^{\alpha}\ket{\boldsymbol{k}+\boldsymbol{G}'}C_{\boldsymbol{k},\boldsymbol{G}'}^{\alpha}=\varepsilon_{\boldsymbol{k}\alpha}C_{\boldsymbol{k},\boldsymbol{G}}^{\alpha}\:.
\end{equation}
Here $\ket{\boldsymbol{k}}$ is shorthand for $\mathcal{V}^{-1/2}e^{i\boldsymbol{k}\cdot\boldsymbol{r}}$, while $H_1^{\alpha}$ denotes the single-particle
Hamiltonian which includes the kinetic energy and the two-dimensional lattice potential (but not the confining potential).

Taking into account a finite number $N_{\boldsymbol{G}}$ of reciprocal-lattice vectors leads to an eigenvalue problem of finite dimensionality.
The latter can be solved numerically to obtain the band dispersion and the Bloch functions \cite{eispackbook}.
In our case, $N_{\boldsymbol{G}}\gtrsim100$ leads to the requisite numerical precision in the diagonalization procedure for all the lattices.

To evaluate the interaction parameters $U_{\alpha\beta}$ [cf. Eq. \eqref{eq:u2d}], we first compute the Wannier functions
\begin{equation}
 \label{eq:wannier}
\widetilde{W}_{\alpha}(\boldsymbol{r})=\frac{1}{\sqrt{N}}\sum_{\boldsymbol{k}}\Psi_{\boldsymbol{k}\alpha}(\boldsymbol{r})\:.
\end{equation}
Because for each value of $\boldsymbol{k}$ the Bloch function is only defined up to a phase factor, the last definition of the Wannier function
is not unique. Therefore, one has to choose a gauge $\Psi_{\boldsymbol{k}\alpha}\rightarrow e^{i\phi(\boldsymbol{k})}\Psi_{\boldsymbol{k}\alpha}$
which leads to localized Wannier functions \cite{kohn_wannier,cloiseaux_wannier}. The substitution \cite{gauge_wannier}
\begin{equation}
\Psi_{\boldsymbol{k}\alpha}(\boldsymbol{r})\rightarrow \exp{\left[-i\operatorname{Im}\ln{\Psi_{\boldsymbol{k}\alpha}(0)}\right]}\Psi_{\boldsymbol{k}\alpha}(\boldsymbol{r})
\end{equation}
gives rise to Bloch functions which have the same phase at $\boldsymbol{r}=0$ for each $\boldsymbol{k}$, leading to Wannier functions which are sufficiently
localized that intersite interactions can be neglected.

\subsection{The optical lattices}
\label{sec:sysmodc}
The optical lattices we investigate are two different 3BLs, one with triangular geometry (3BTL) and one with square geometry (3BSL), and the four-beam square lattice (4BSL).
Hereafter, the four-beam and three-beam lattice potentials are respectively labeled by the superscripts 4B and 3B:
\begin{eqnarray}
\nonumber
 V^{4B}(\boldsymbol{r})&=\dfrac{V_0}{2}\Big[&\cos(2k_Lx)+\cos(2k_Ly)\Big], \\ \nonumber
 V^{3B}(\boldsymbol{r})&=\dfrac{V_0}{2}\Big[&\cos(\boldsymbol{b}_1\cdot\boldsymbol{r})+\cos(\boldsymbol{b}_2\cdot\boldsymbol{r})\\ \label{eq:potts}
 &&+\cos([\boldsymbol{b}_1+\boldsymbol{b}_2]\cdot\boldsymbol{r})\Big].
\end{eqnarray}
Here $\boldsymbol{b}_i\equiv\boldsymbol{k}_i-\boldsymbol{k}_{i+1}$ are the reciprocal-lattice vectors, where $\boldsymbol{k}_i$ denotes the wave-vector of the $i$-th laser beam with magnitude $k_L=2\pi/\lambda_L$.
To produce the desired lattice geometries, the lasers need to be red detuned [$V_0<0$ in Eq. \eqref{eq:potts}].
The magnitude of $V_0$ is in principle dependent on the atomic properties of the Bose components. In this work, however, we consider systems
where $V_0$ is approximately equal for both species. This is not an unreasonable assumption since such systems are experimentally accessible \cite{bboptlatt}.

\begin{figure}[h]
\includegraphics[width=.5\textwidth]{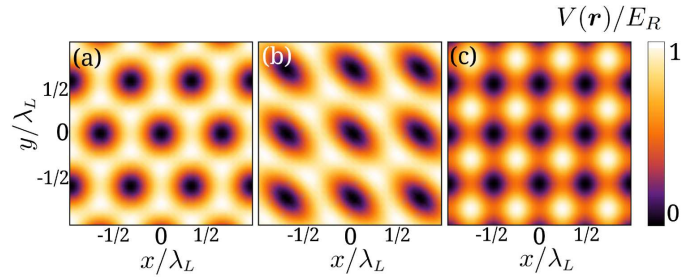}
%%[trim=40mm 1mm 40mm 1mm,clip,width=1.75cm]
\caption{(Color online) Optical-lattice potentials in units of the recoil energy $E_R=k^2/2m_A$ with their minima set equal to zero. (a) 3BTL, (b) 3BSL,
(c) 4BSL. For these plots $V_0=-E_R/2$.}
\label{fig:potts}
\end{figure}

The 4BSL is created by four laser beams of equal intensity and polarization enclosing mutual angles of $\pi/2$ and has a lattice period of $\lambda_L/2$. For the 3BSL, the
three laser beams enclose the angles $\pi/2$, $\pi/2$, and $\pi$ resulting in a lattice spacing of $\lambda_L/\sqrt{2}$, while for the 3BTL all the enclosing angles are $2\pi/3$ and the lattice spacing
is $2\lambda_L/3$. The different lattice geometries are shown in Fig. \ref{fig:potts}.

Straightforward evaluation of the matrix elements of $H_1^{\alpha}$ in the plane-wave basis yields
\begin{widetext}
\begin{eqnarray}
\nonumber
&&\bra{\boldsymbol{k}+\boldsymbol{G}_{l,n}}H_1^{\alpha}\ket{\boldsymbol{k}+\boldsymbol{G}_{l,n}}=\frac{1}{2m_{\alpha}}\abs{\boldsymbol{k}+\boldsymbol{G}_{l,n}}^2,\\\label{eq:matrixelements}
&&\bra{\boldsymbol{k}+\boldsymbol{G}_{l_1,n_1}}V^{4B}(\boldsymbol{r})\ket{\boldsymbol{k}+\boldsymbol{G}_{l_2,n_2}}
=\frac{V_0}{4}\Big(\delta_{l_1,l_2\pm1}\delta_{n_1,n_2}+\delta_{l_1,l_2}\delta_{n_1,n_2\pm1}\Big),\label{eq:offdiagonalfour}\\\nonumber
&& \bra{\boldsymbol{k}+\boldsymbol{G}_{l_1,n_1}}V^{3B}(\boldsymbol{r})\ket{\boldsymbol{k}+\boldsymbol{G}_{l_2,n_2}}
=\frac{V_0}{4}\Big(\delta_{l_1,l_2\pm1}\delta_{n_1,n_2}+\delta_{l_1,l_2}\delta_{n_1,n_2\pm1}+\delta_{l_1,l_2\pm1}\delta_{n_1,n_2\pm1}\Big),
\end{eqnarray}
\end{widetext}
where $\boldsymbol{G}_{l,n}\equiv l\boldsymbol{b}_1+n\boldsymbol{b}_2$ ($l,n$ are integers) are the reciprocal-lattice vectors.
The diagonal matrix elements are independent of the lattice potential, while
the off-diagonal ones only depend on the potential and are equal for all 3BLs.

\section{Derivation of the superfluid-drag coefficient}
\label{sec:drag}
The derivation treated in this section generalizes Ref.~\onlinecite{linder_drag}; details are therefore omitted.
Considering the case where the ground state ($\boldsymbol{k}=0$) is macroscopically occupied, the corresponding boson
creation and annihilation operators commute to a very good approximation. Thus we can replace them by c-numbers \cite{pethicksmith}:
\begin{eqnarray}
 \label{eq:bogoliubov}
 \nonumber
 \langle b_{0\alpha}b^{\dag}_{0\alpha}\rangle&=&\langle b^{\dag}_{0\alpha}b_{0\alpha}\rangle+1\approx N_{0\alpha}=\langle b^{\dag}_{0\alpha}b_{0\alpha}\rangle\:,\\
 b_{0\alpha}&\approx&b^{\dag}_{0\alpha}\approx \sqrt{N_{0\alpha}}\:.
\end{eqnarray}
We then neglect all the terms of higher order than bilinear in the creation/annihilation operators of $\boldsymbol{k}\neq0$ states.
The resulting bilinear Hamiltonian can be diagonalized by following the procedure outlined in Ref.~\onlinecite{Tsallis:78}. In terms of the Bogoliubov quasiparticle operators $\beta_{\boldsymbol{k}\sigma}$,
one obtains
\begin{equation}
 \label{eq:hamdiag}
H=H_0-\frac{1}{2}\sideset{}{'}\sum_{\boldsymbol{k},\alpha}E_{\boldsymbol{k}}^{\alpha}+\sideset{}{'}\sum_{\boldsymbol{k},\sigma}\mathcal{E}_{\boldsymbol{k}\sigma}
\left(\beta^{\dag}_{\boldsymbol{k}\sigma}\beta_{\boldsymbol{k}\sigma}+\frac{1}{2}\right),
\end{equation}
where the primed sum runs over all $\boldsymbol{k}\neq0$ states and the two-branch ($\sigma=\pm$) excitation spectrum is given by
\begin{widetext}
\begin{equation}
 \label{eq:spectrum0}
\mathcal{E}_{\boldsymbol{k}\sigma}=\frac{1}{\sqrt{2}}\left\{\epsilon_{\boldsymbol{k}}^{A}\left(\epsilon_{\boldsymbol{k}}^{A}+2F_A\right)+
\epsilon_{\boldsymbol{k}}^{B}\left(\epsilon_{\boldsymbol{k}}^{B}+2F_B\right)+\sigma\sqrt{\left[\epsilon_{\boldsymbol{k}}^{A}\left(\epsilon_{\boldsymbol{k}}^{A}+2F_A\right)-
\epsilon_{\boldsymbol{k}}^{B}\left(\epsilon_{\boldsymbol{k}}^{B}+2F_B\right)\right]^2+16F_{AB}^2\epsilon_{\boldsymbol{k}}^{A}\epsilon_{\boldsymbol{k}}^{B}}\right\}^{1/2}.
\end{equation}
\end{widetext}
Here $F_{\alpha}=n_{\alpha}U_{\alpha\alpha},\:\: F_{AB}=\sqrt{n_An_B}\;U_{AB},\:\:\:
\epsilon_{\boldsymbol{k}}^{\alpha}=\varepsilon_{\boldsymbol{k}\alpha}-\varepsilon_{0\alpha},$ and $E_{\boldsymbol{k}}^{\alpha}=\epsilon_{\boldsymbol{k}}^{\alpha}+F_{\alpha}$;
$n_{\alpha}\equiv N_{\alpha}/N$ is the particle density of component $\alpha$.
It is easy to see that this spectrum is gapless. Since the Bogoliubov approximation makes the assumption of a small condensate depletion, only the superfluid phase can be described
by the obtained quantities \cite{oosten:01}.

For small superfluid velocities $\boldsymbol{v}_{\alpha}$, the free energy of a two-component Bose-gas can be expanded as \cite{superfluid_drag}
\begin{equation}
 \label{eq:free_energy}
F=F_0+\frac{\mathcal{V}}{2}\left[\rho_{A}^s\boldsymbol{v}_A^2+\rho_{B}^s\boldsymbol{v}_B^2-\rho_d\left(\boldsymbol{v}_A-\boldsymbol{v}_B\right)^2\right],
\end{equation}
where $\rho_{\alpha}^s$ is the superfluid density of component $\alpha$, while $F_0$ denotes the terms independent of the superfluid velocities.
At zero temperature, the free energy is equal to the expectation value of the Hamiltonian with no quasiparticles excited
\begin{equation}
 \label{eq:free_energy2}
F_{T=0}=\langle H_0\rangle+\frac{1}{2}\sideset{}{'}\sum_{\boldsymbol{k}}\left(\mathcal{E}_{\boldsymbol{k}+}+\mathcal{E}_{\boldsymbol{k}-}-E_{\boldsymbol{k}}^A-E_{\boldsymbol{k}}^B\right).
\end{equation}

The superfluid density of a one-component system is determined by its response to an externally induced superfluid velocity \cite{yukalov}.
To determine the drag, the corresponding two-component generalization of this procedure reads \cite{linder_drag}
%Introducing small superfluid velocities in the individual boson components leads to a change in particle momentum and energy
\begin{eqnarray}
\nonumber
 \boldsymbol{k}&\rightarrow&\boldsymbol{k}-m_{\alpha}\boldsymbol{v}_{\alpha}\:,\\ \label{eq:transformation}
 \epsilon_{\boldsymbol{k}}^{\alpha}&\rightarrow&\epsilon_{\boldsymbol{k}}^{\alpha}-m_{\alpha}\boldsymbol{v}_{\alpha}\cdot\nabla_{\boldsymbol{k}}\epsilon_{\boldsymbol{k}}^{\alpha}+\mathcal{O}(v_{\alpha}^2)\:.
\end{eqnarray}
Using this transformation, the superfluid-drag coefficient can be found by expanding the free energy in the superfluid velocities. Considering the case of parallel superfluid flows,
we find
\begin{equation}
 \label{eq:rhod}
\rho_d=\frac{1}{\mathcal{V}}\sideset{}{'}\sum_{\boldsymbol{k}}\frac{2m_{A}m_{B}F_{AB}^2\epsilon_{\boldsymbol{k}}^{A}\epsilon_{\boldsymbol{k}}^{B}}
{\mathcal{E}_{\boldsymbol{k}+}\mathcal{E}_{\boldsymbol{k}-}\left(\mathcal{E}_{\boldsymbol{k}+}+\mathcal{E}_{\boldsymbol{k}-}\right)^3}
\left(\partial_{k_u}\epsilon_{\boldsymbol{k}}^{A}\right)\left(\partial_{k_u}\epsilon_{\boldsymbol{k}}^{B}\right),
\end{equation}
where $\boldsymbol{\hat{u}}$ denotes the direction of the superfluid flow and
$\partial_{k_u}\equiv\boldsymbol{\hat{u}}\cdot\boldsymbol{\nabla}_{\boldsymbol{k}}$ stands for the corresponding directional derivative. 
Note that the drag is always positive and independent of the sign of the interspecies interaction. The last expression holds for an arbitrary lattice.

\section{Results and discussion}
\label{sec:results}
\begin{figure}[b]
\includegraphics[width=.5\textwidth]{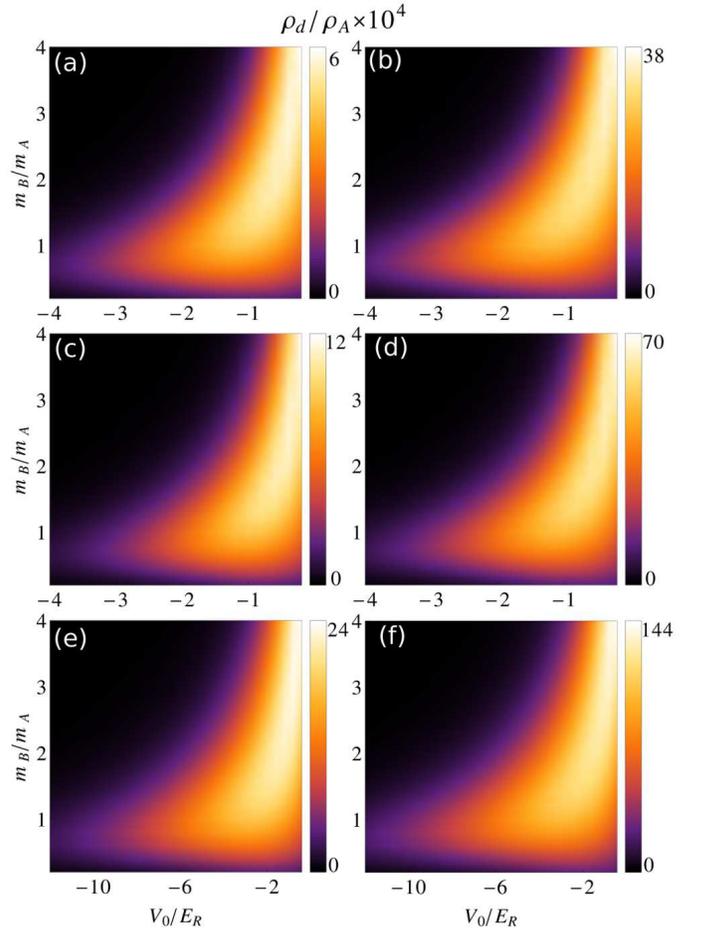}
%%[trim=40mm 1mm 40mm 1mm,clip,width=1.75cm]
\caption{(Color online) Superfluid drag for different lattices and interaction strengths: (a) and (b) three-beam triangular lattice (3BTL),
(c) and (d) three-beam square lattice (3BSL), (e) and (f) four-beam square lattice (4BSL). The left column [(a), (c), (e)] corresponds to weak interspecies interactions
$a_{AB}=30\:a_0$, the right column [(b), (d), (f)] to strong ones $a_{AB}=64\:a_0$. Note the different scales for the color scheme.}
\label{fig:results}
\end{figure}
In the following, we present our findings for the superfluid drag in different lattice geometries, for weak and strong interspecies scattering.
To obtain a dimensionless quantity, we normalize the superfluid drag by $\rho_A=N_Am_A/\mathcal{V}$.
Superfluidity at all lattice depths is assured by choosing an incommensurate filling $n_A=n_B=\sqrt{2}$ \cite{Chen+Wu:03}. The intraspecies scattering lengths are set to $a_{AA}=100\:a_0$ ($a_0$ is the Bohr radius)
and $a_{BB}=65\:a_0$. We choose the superfluid flows to be codirected in the $x$-direction defined in Fig. \ref{fig:potts} for all the lattices.
For the laser wavelength, which determines the lattice spacing, we choose $\lambda_L=1064\:$nm.
The results obtained for the superfluid drag are shown in Fig. \ref{fig:results}. Figures \ref{fig:results}(b), (d), (f) correspond to an interspecies interaction just below the value where phase separation occurs
[i.e., where the excitation spectrum in Eq. \eqref{eq:spectrum0} becomes imaginary].

\begin{figure}[t]
\includegraphics[width=.45\textwidth]{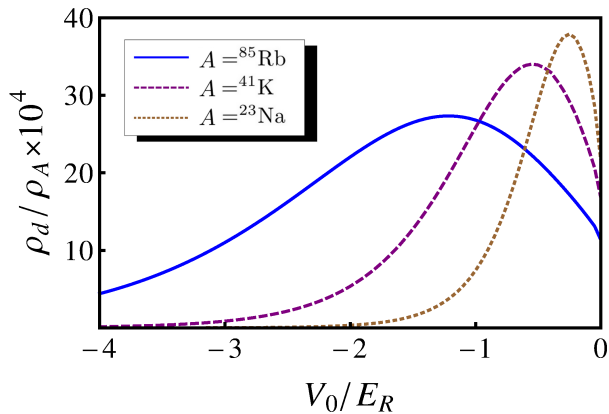}
%%[trim=40mm 1mm 40mm 1mm,clip,width=1.75cm]
\caption{(Color online) Superfluid drag in a 3BTL as a function of the lattice depth $V_0$ for fixed mass ratios corresponding to the mixtures $^{87}$Rb-$^{85}$Rb ($m_B/m_A\approx1$, solid), $^{87}$Rb-$^{41}$K
($m_B/m_A\approx2.2$, dashed) and $^{87}$Rb-$^{23}$Na ($m_B/m_A\approx3.8$, dotted). Component $B$ corresponds to $^{87}$Rb in all three cases. The interspecies scattering length is set to $a_{AB}=64\:a_0$.}
\label{fig:oned}
\end{figure}

\begin{figure}[b]
\includegraphics[width=.4\textwidth]{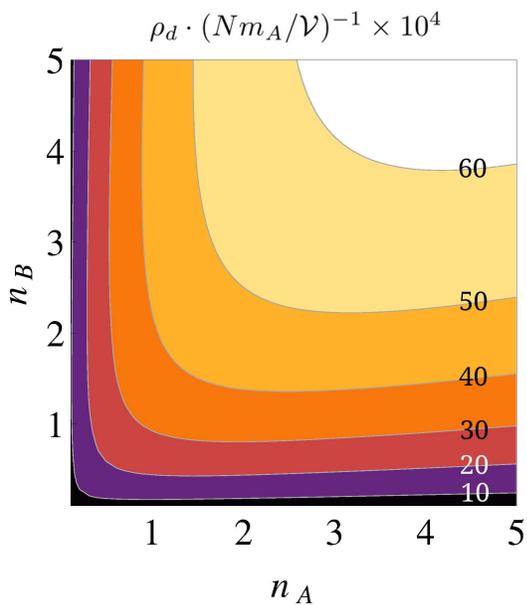}
%%[trim=40mm 1mm 40mm 1mm,clip,width=1.75cm]
\caption{(Color online) Superfluid drag in a 3BTL (\textit{not} normalized with $n_A$) as a function of the particle densities for $A=^{85}$Rb and a lattice depth $V_0=-1.2\:E_R$ which maximizes the drag
for this mass ratio (see Fig. \ref{fig:oned}). The other parameter values are the same as in Fig. \ref{fig:results} (b). Component $B$ corresponds to the more weakly interacting species.}
\label{fig:ndep}
\end{figure}

Keeping the intraspecies interactions constant, the drag effect is enhanced significantly with increasing interspecies interactions. In contrast to the results of Ref.~\onlinecite{linder_drag},
however, we find that the mass ratio which maximizes the drag depends on the lattice depth, varying significantly from unity as one goes to shallower lattices ($|V_0|\lesssim1\:E_R$).
For clarity, cuts of Fig. \ref{fig:results}(b) for experimentally relevant mass ratios are presented in Fig. \ref{fig:oned}.

Although the magnitude of the superfluid drag varies strongly with the lattice geometry, the qualitative behavior as a function of the mass ratio and the lattice depth is the same:
for fixed mass ratio, the drag increases upon raising $|V_0|$ from zero, reaches its maximum and then decreases. Quantitatively, this behavior is strongly dependent on the mass ratio.

The lattice-depth dependence of the drag can be ascribed to a competition between the interaction and kinetic energies. As $|V_0|$ increases, the interactions become stronger and the drag, which depends
quadratically on $F_{AB}$, is enhanced. At the same time, the bandwidth, incorporated in $\rho_d$ through the derivative of the single-particle dispersion, decreases with increasing lattice depth.
This effect is strongly dependent on the particle masses involved.
A similar behavior can be observed upon increasing $m_B$ for a fixed lattice depth. For a small mass ratio the drag increases,
because it is linear in $m_B$, before it starts to decay due to the decrease in kinetic energy.

Comparing the 3BSL with the 4BSL, one notices that in the 4BSL the superfluid drag is much larger over a broader range of lattice depths. This is consistent with the fact that
the hopping amplitudes are smaller and the SF-MI transition sets in at much shallower lattices for the 3BLs \cite{blakie:04}. A possible explanation as to why the drag is stronger in the 3BSL than in the 3BTL can be given
if one envisions the drag as being mediated by component $A$ particles, dressed by a cloud of component $B$ particles and vice versa. In this case one could argue that the higher coordination number of the 3BTL leads to
more events whereby one of the particles within the cloud splits off the dressed particle.

In addition to the above findings, we stress that for a fixed mass ratio, the lattice depth $V_0$ which maximizes the drag seems to be independent of the particle densities $n_{\alpha}$. On the other hand, the magnitude of 
$\rho_d$ depends on the particle densities as shown in Fig. \ref{fig:ndep}. At fixed $V_0$, this dependence itself varies with the mass ratio. Comparing the density dependence at
$V_0$ which maximizes the drag for the respective mass ratio, we find that it hardly changes at all.

Finally, we note that for two components with the same scattering length the drag always increases upon increasing the particle number (data not shown). For two species with different
scattering lengths on the other hand, increasing the number of the more strongly interacting particles can actually lead to a decrease in the superfluid drag (see Fig. \ref{fig:ndep}).

\section{Conclusions}
\label{sec:conclusion}
We have studied the drag between components of a two-species BEC in optical lattices. To assure that our results are valid even for shallow optical lattices, we did not make use of the tight-binding
approximation but used a numerically exact approach instead. Consequently, we generalized the previously derived expression for the superfluid-drag coefficient to arbitrary lattice geometries.
To clarify the dependence of the superfluid drag on the lattice geometry, we have presented results for rectangular and non-rectangular lattices with separable and non-separable potentials.

We have demonstrated a non-monotonic dependence of the drag on the lattice depth that results from the competition between two effects:
the drag increases with the interspecies interaction strength and is reduced upon
decreasing the kinetic energy. While the qualitative behavior of the drag is the same for all the lattice geometries studied, its quantitative properties, such as the magnitude, differ from one lattice to another.

Our study will hopefully motivate drag experiments with ultracold atoms in optical lattices.

\begin{acknowledgments}
This work was financially supported by the Swiss SNF,
the NCCR Nanoscience, and the NCCR Quantum Science
and Technology. 
\end{acknowledgments}

% \bibliographystyle{apsrev}
% 	\bibliography{paper.bib}

\end{document}